%% file: main.tex
\documentclass[10pt,conference]{IEEEtran}

\usepackage{cite}

\usepackage{graphicx}
\usepackage{todonotes}
\usepackage{braket}
\usepackage[nohyperlinks]{acronym}
\usepackage{hyperref}
\usepackage{amsmath, amssymb}
\usepackage{tikz}
\usetikzlibrary{calc}

\begin{document}

\input{acronyms.tex}

\title{Sampling problems on a Quantum Computer}

\author{\IEEEauthorblockN{Maximilian Balthasar Mansky\IEEEauthorrefmark{1}, Jonas Nüßlein\IEEEauthorrefmark{1}, David Bucher\IEEEauthorrefmark{2}, Daniëlle Schuman\IEEEauthorrefmark{1}, Sebastian Zielinski\IEEEauthorrefmark{1}, \\
and Claudia Linnhoff-Popien\IEEEauthorrefmark{1}}
\IEEEauthorblockA{\IEEEauthorrefmark{1}LMU Munich\\
Munich, Germany\\
Email: maximilian-balthasar.mansky@ifi.lmu.de}
\IEEEauthorblockA{\IEEEauthorrefmark{2}Aqarios GmbH\\Munich, Germany}}

 




\date{\today}

\maketitle

\begin{abstract}
Due to the advances in the manufacturing of quantum hardware in the recent years, significant research efforts have been directed towards employing quantum methods to solving problems in various areas of interest. Thus a plethora of novel quantum methods have been developed in recent years. In this paper, we provide a survey of quantum sampling methods alongside needed theory and applications of those sampling methods as a starting point for research in this area. This work focuses in particular on Gaussian Boson sampling, quantum Monte Carlo methods, quantum variational Monte Carlo, quantum Boltzmann Machines and quantum Bayesian networks. We strive to provide a self-contained overview over the mathematical background, technical feasibility, applicability for other problems and point out potential areas of future research.
\end{abstract}

\section{Introduction}\label{sec:Introduction}

Sampling from a population is a well established way to learn about the structure of large data or to learn about a distribution of properties in spaces that are too large for an exhaustive examination. Sampling is the act of drawing samples from some unknown distribution with the goal of learning about the underlying distribution by using statistical reasoning. One samples to generate a smaller population from a distribution because querying all possible elements is prohibitively expensive. For many problems sampling provides a reasonable approximation of the total distribution and is an important approximative method. Sampling is used to build an understanding of complex systems with limited resources and therefore an important tool in many areas of science. \cite{witte_statistics_2017}. 


For reliable statistical reasoning, one needs a sufficient amount of samples. The number of samples depends on the problem at hand and the applicable statistical rigour and is an important consideration when using a sampling approach. Samples also need to be independent and identically distributed (i.i.d.), meaning that subsequent samples should have no relationship with each other and should not affect each other's result and that they need to come from the same total distribution. In turn, this means that sampling processes can be easily parallelized for fast querying.

These benefits of sampling approaches to problem solving, namely fast approximative solutions to complex systems explain their ubiquity and applicability across a broad range of subjects \cite{agresti_art_2007, thompson_sampling_2012}. There is continuous effort to improve current theory and provide better approaches to creating independent samples for individual problems.

A novel approach in generating samples is the use of a quantum computer. Quantum computing is a new computational paradigm with promises of significant computational speedup \cite{nielsen_quantum_2010}. The technology is structurally different from classical computing and relies on the effects of quantum mechanics to process information. Instead of representing information through a binary encoding, quantum computing relies on the superposition of states for encoding. The states have a physical representation depending on the underlying hardware and are typically different energy levels in a quantum-mechanical system. Superposition means that the state is composed from a complex-valued superposition of base states. Information can be changed on the quantum computer through operations called gates and can be retrieved for further classical processing and interpretation with a measurement.

The two approaches, quantum computing and sampling from the unknown, can be combined to shed light on distributions within quantum-mechanical systems that are otherwise difficult to calculate or to model classically. In the context of sampling, this means obtaining a sample via classical simulation is difficult, but drawing a sample via a quantum computer is easy. Many of the physical systems that one is interested in form a high-dimensional space whose structure is difficult to model. Even simple composite two-state systems suffer from a combinatorial explosion in their complete state representation, where the number of dimensions scales as $\mathcal{O}(2^n)$, with $n$ the number of qubits. Since the system size of a quantum computer scales in the same way, a quantum computer is an appropriate tool to simulate these systems. This is reflected in the approaches discussed in this paper, drawing inspiration from physical systems.

Classical distributions can also be modeled on a quantum computer, taking advantage of the larger representation space in order to represent more complex distributions and then draw samples from them. The expressiveness of these systems can be higher than an equivalent classical formulation, again due to the larger available state space. 


The topic of quantum-based sampling has received significant attention over the past few years, with experimental breakthroughs in the size of the experiments \cite{madsen_quantum_2022}.

In this work we introduce the state of the art of quantum sampling techniques and applications and provide an overview of their technical feasibility, current technical state and applicability as a solution to other problems. We present the topic as a self-contained as possible to allow a fast translation towards a solution. 

In the remainder of the paper we provide a background on sampling as an explicit approach on classical and quantum systems. In section \ref{sec:examples} we introduce all current approaches to quantum-based sampling, including their mathematical basis, technological readiness, open questions and applicability as a solution path. Lastly we provide a discussion of the presented methods.
 
\section{Background on sampling techniques}

A sampling algorithm can be imagined as a machine that transforms uniform random bits into non-uniformly distributed random bits \cite{lund_quantum_2017} In the context of sampling from a population, it means taking independent and identical samples from the distribution. This simple structure is a starting point into questions of statistics, probability modeling, conditional inference and much more. The sampling algorithm that transforms the bit strings is often opaque to the questions, if known at all.

\subsection{Sampling on a classical system}

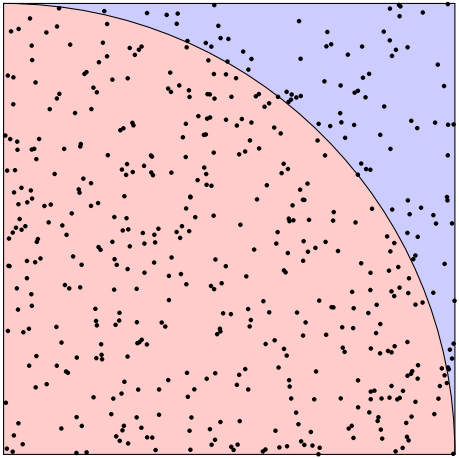
\begin{figure}
    \centering
    \begin{tikzpicture}[scale=6]
        \fill[blue!20] (0,0) rectangle (1,1);
        \fill[red!20] (0,0) -- (1,0) arc [start angle = 0, end angle = 90, radius = 1] --cycle;
        \draw (0,0) rectangle (1,1);
        \draw (1,0) arc [start angle = 0, end angle= 90, radius = 1];
        \foreach \i in {1,...,500}
            {\fill (rnd, rnd) circle (.005);            
            }
    \end{tikzpicture}
    \caption{Visualization of a sampling process to determine the value of $\pi$ via a sampling process. Black dots represent random $(x,y)$ positions. If the distance to the origin is smaller or equal to $1$ (red area), it is counted towards the red bin, otherwise it is only counted towards the total number of samples. The fraction of the count of elements in the red bin to the total number of samples multiplied by $4$ approximates $\pi$.}
    \label{fig:pi}
\end{figure}

A sampling algorithm turns a source of uniform randomness into a non-uniform one. Running the algorithm many times generates a statistical sample with meaningful insights into the underlying problem. A classic example is sampling $\pi$ by generating uniform random $x, y$ positions in the unit square. Here, the sampling algorithm takes uniform randomness -- the position -- and assigns a new value based on the points' distance from the origin. After generating many points, the fraction of points within unit distance of the origin provide an estimate of the value of $\pi$. This is visualized in Figure \ref{fig:pi}.

\subsection{Sampling on quantum circuits}

A quantum computer can similarly be used for a) performing calculations and b) providing a translation between uniform randomness and a biased distribution. The necessary randomness is however not easy to control, despite the fact that a quantum computer is fundamentally a probabilistic machine \cite{nielsen_quantum_2010}. The obvious quantum circuit for generating randomness, applying a Hadamard gate to each available qubit in the $|0\rangle$ state, results in uniform independent random \emph{qubits} when measuring in the same $\langle0|$ basis. In order to represent any uniformly random \emph{state} of the Hilbert space, one needs an exponentially deep circuit in the number of qubits \cite{haferkamp_linear_2022}. This is in stark contrast to the classical world, where the randomness of a bit string is fundamentally the same as the randomness of an equivalently sized group of individually sampled bits \cite{herrero-collantes_quantum_2017}.

Suitable construction of the quantum circuit can also generate an appropriate sampling statistic without a source of randomness. This makes use of the probabilistic nature of a quantum computer, where a measured mixed state returns a probabilistic value based on the measurement basis. Mixed states form a statistical ensemble that can be expressed through a density matrix $\rho$ \cite{nielsen_quantum_2010}.

For this discussion, we assume a perfect quantum computer without extrinsic noise. The noise of current imperfect quantum computers is an unwanted source of randomness and will in most cases bias the desired calculation.

\section{Examples of sampling problems}\label{sec:examples}

We provide a thorough coverage of different quantum computing approaches to sampling problems. We provide a self-contained background on the mathematical basis. Technical feasibility on current hardware is explored as well, especially for the cases where an implementation has already been achieved. We also highlight the applicability as a solution path to other areas of science and take note of open questions for future work. 

We start with Gaussian Boson Sampling, an experiment on current hardware that explicitly samples on an unknown distribution and is very time-consuming to model classically. Quantum-Enhanced Markov Chain Monte-Carlo expands the classical MCMC structure to quantum computers with promises of significant speed-up when investigating quantum systems. Variational Monte Carlo is especially applicable to material science and chemistry. Quantum Boltzmann Machines are a sampling-based machine learning model that take advantage of the quantum computing structure for faster and easier sampling from arbitrary distributions. Lastly, quantum Bayesian Networks are a direct translation of classical models for chained probabilities. There are indications that the quantum version has significantly higher expressive power.

\subsection{Gaussian Boson Sampling}

Boson Sampling is a simplified, non-universal model of quantum computation first introduced by Aaronson and Arkipov \cite{AA} in which $n$ Bosons, originally in an input arrangement $k$, are scattered by a passive, linear unitary transformation $U$ into $m \gg n$ output modes. The Boson Sampling problem consists of producing a fair sample of the output probability distribution $P (l|k, U )$, where $l$ is the output arrangement \cite{Zhong, Lund}. Aaronson and Arkipov argue that the existence of an efficient classical algorithm which accomplishes this given a random transformation $U$ implies the ability to estimate the permanent of an arbitrary complex valued matrix, a problem lying in the class $\#P$ \cite{AA}. This means that the problem is in fact hard for classical computers to solve and provides an argument for the superiority of quantum computers over classical ones, as the Boson Sampling problem can be solved efficiently by the former, as well as evidence against the Church-Turing thesis \cite{Lund}.

The primary hurdle in implementing Boson Sampling experimentally lies in the fact that currently available single photon sources are spontaneous, meaning that the cost of producing an input state with exactly $n$ photons grows exponentially in $n$. To combat this issue, Lund et al. \cite{Lund} suggest using Gaussian states, which can be produced deterministically with high purity. They describe a \textit{Gaussian Boson Sampler}, 
a quantum optical processor consisting of 2-mode squeezed input states and a non-adaptive linear optical network, which produces photon number counting statistics as its output. They argue that in one particular case, namely in the context of the generalized Boson Sampling problem, such a device can efficiently sample distributions which are hard to sample for classical counterparts. Furthermore, Lund et al. contend that approximate Boson Sampling is also a hard problem, even in the generalized case \cite{Lund}.

In \cite{Hamilton}, Hamilton et al. formally introduce Gaussian Boson Sampling (GBS), which, unlike previous protocols involving Gaussian states, takes full advantage of the Gaussian nature of the states. In the GBS setup, Single Mode Squeezed States (SMSS) enter a linear interferometer, and the output patterns are sampled in the photon number basis. They show that the probability of measuring a specific output distribution of a Gaussian input state is related to the \textit{hafnian}, a matrix function more general than the permanent which resides in the $\#P$ complexity class \cite{Hamilton, Kruse}. With this result, Hamilton et al. prove that the GBS protocol resides in $\#P$ along with the approximate sampling problem. This protocol differs above all due to the fact that the sampling matrix describes not only the action of the interferometer, but also the shape of the Gaussian input state. This implies that a coherent superposition of all $n$-photon patterns from the input can be used and no exact input pattern must be heralded as in other protocols. As such, GBS increases photon generation probability relative to standard boson sampling protocols which use single photon Fock states \cite{Zhong}. Furthermore, GBS reduces the size of the sampling space by a factor of $\binom{N^2}{N}$ compared to Scattershot Boson sampling, thereby significantly advancing experimental possibilities. While the classification of the Boson Sampling problem with Gaussian states has not definitively been assigned a complexity class, it has been shown that the special case of sampling from a multimode thermal state resides in $BPP^{NP}$ \cite{Hamilton}. In \cite{Kruse}, Kruse et al. built upon the work of Hamilton by adjusting the protocol to account for displaced squeezed states and higher-order photon number contributions. 

Since the conception of GBS, a number of experimental implementations have advanced the study of the protocol. Most notably, Zhong et al. \cite{Zhong20} used a photonic quantum computer, Jiuzhang, to execute the GBS protocol with 50 indistinguishable single-mode squeezed states and a 100-mode ultralow-loss interferometer with full connectivity and sampled the output using 100 high-efficiency single-photon detectors. Jiuzhang has a 76-photon coincidence, with an output state space dimension of $10^{30}$ and outpaced classical state-of-the-art simulation on supercomputers by a factor of $~10^{14}$. 

Significant progress has also been achieved in the classical simulation of GBS. Bulmer et al. \cite{Bulmer} present a classical GBS simulation method using threshold detectors, which demonstrates a nine-orders of magnitude speedup over previous classical algorithms that employ photon number–resolving detectors. The novel GBS simulation using threshold detectors was applied to two separate sampling algorithms, a probability chain rule method and Metropolis independence sampling, and was able to simulate the GBS protocol with up to 92 photons and 100 modes, reducing computation time from 600 million years to a matter of months, a nine-orders of magnitude improvement. However, such an approach only proves useful for verification purposes, as state-of-the-art GBS setups, such as Jiuzhang, require only minutes for the same computation. 

\input{monte_carlo}\label{sec:monte-carlo}


\subsection{Bayesian networks}

Bayesian networks are used to represent chained probabilities \cite{stephenson_introduction_2000}. They are used to represent certainty about knowledge and to model information insecurity. As they can be used to calculate difficult probability distributions, they have found wide application across a number of areas, including agriculture \cite{drury_survey_2017}, data mining \cite{heckerman_bayesian_1997}, meteorology \cite{cano_applications_2004}, software quality control \cite{tosun_systematic_2017} and many other areas with distributed influence factors.

Mathematically, the expected probability over each edge of the network is given by
\begin{equation}
    p(x_1, \ldots, x_n) = \prod_{i=1}^n p(x_i | \text{parents of }x_i)
\end{equation}
This structure can be modelled through directed acyclic graphs (DAG), where each probability $p_i$ is assigned to a node and the parental relations are mapped to edges between the nodes. The directionality of the edges is a consequence of the parent-child relationship between the probabilities, which also prohibits cyclic relationships inside the network. Finding the correct Bayesian network for a given set of data is generally difficult (NP-complete in the general case \cite{chickering_learning_1996}), but there are a number of approximate algorithms and heuristics that improve training time \cite{koller_probabilistic_2009}.

These networks can in principle be mapped to a quantum computer \cite{low_quantum_2014, gao_enhancing_2022}. There is a direct mapping between nodes and edges of the network and the operators on a quantum computer \cite[fig: 9]{gao_enhancing_2022}. Using $k$-controlled unitaries, where the controls to the $i$th qubit correspond to the edges on the equivalent graph formulation, a state is prepared such that $|\langle x_i | U(\text{parents of }x_i)|x_i\rangle|^2 = p(x_i| \text{parents of } x_i)$. The correspondence between the Bayesian network graph and its the quantum circuit is shown in figure \ref{fig:bayesian-network}.

\begin{figure}
    \centering
    \begin{tikzpicture}
\draw (0,1) node (q1) {}
	 (0,4) node (q4) {}
	 (0, 2) node (q2) {}
	 (1, 3) node (q3) {}
	 (2, 5) node (q5) {};
\foreach \i in {1,...,5} {
	\node at (q\i) [circle, fill=blue!20, draw=blue] (q\i) {$q_\i$};
	};
\draw[->] (q2) -- (q3);
\draw[->] (q4) -- (q3);
\draw[->] (q4) -- (q5);
\draw[->] (q3) -- (q5);
\draw[->] (q1) to[bend right] (q5);
\begin{scope}[xshift=4cm]
\foreach \i in {1, ...,5} {
	\draw (0, \i) node (start\i) {} --+ (3,0);
	\node at (start\i) [left] {$|\psi_\i\rangle$};
	\draw (3, \i) node[rectangle, draw=black, fill=white] {
		\begin{tikzpicture}
			\draw (60:.5) arc [start angle =60, end angle = 120, radius = .5];
			\draw (80:.4) -- (70:.6);
		\end{tikzpicture}
	};
};
\filldraw (1, 2) circle (.05) -- (1, 4) circle (.05);
\draw (1, 3) node[rectangle, draw=black, fill=white] {$U_3$};
\filldraw (2,1) circle (.05) -- (2,3) circle (.05) -- (2,4) circle (.05) -- (2,5);
\draw (2,5) node[rectangle, draw=black, fill=white] {$U_5$};
\end{scope}
\end{tikzpicture}
    \caption{A Bayesian network and its corresponding quantum circuit. Each connection in the graph is translated to a control wire. The influence of each parent node to its children is modeled through unitary gates on the corresponding qubit. Adapted from \cite{gao_enhancing_2022}}
    \label{fig:bayesian-network}
\end{figure}

The approach differs from the ones discussed in section \ref{sec:mc-enhanced} in that rather than using a learning framework where the system is changed to approximate some desired solution, the Bayesian networks are constructed rather schematically as a translation of a known system or an ansatz expression of the DAG. 

Structurally, these models should be expandable to more general sequential quantum models. The construction on the quantum computer may be able to model any sequential structure. Sequential and ordered structures are ubiquitous across applications, from biological systems to natural language processing.

Quantum natural language processing is already in development \cite{zeng_quantum_2016, meichanetzidis_grammar-aware_2023}. Its structure is at first dissimilar to the chained probability approach presented here but in practice follows much the same structure of distributed meanings translated to quantum circuits that are then chained together to form a quantum computation \cite{coecke_mathematical_2010}. Bridging between the two approaches can bring a meaningful contribution in understanding for both fields. 

Biological systems, especially where they model complex interactions, are another field of application where Bayesian networks find wide application \cite{needham_inference_2006, liu_application_2010, schaapveld_bayesian_2019}. Leveraging the additional structure of quantum Bayesian networks can lead to additional insights into the structure of relationships between different influence factors. This can be used to model disease vectors \cite{lau_bayesian_2016}, disease recognition \cite{seixas_bayesian_2014}, neuroscientific discoveries \cite{bielza_bayesian_2014} or the analysis of genomes \cite{sebastiani_bayesian_2005}. 

Beyond static Bayesian networks, dynamic Bayesian networks that reflect a change of values and possibly structure of the Bayesian network \cite{murphy_dynamic_2002, ghahramani_learning_2006, song_time-varying_2009} may also be applicable to quantum ansatz presented here. Their expressive power is significantly higher and can be suited to the higher dimensional space of quantum computers.

\section{Discussion}


In this paper, we review sampling applications as an important application area for quantum computing.
We start by defining sampling as a means of approximating complicated distributions with queries from that distribution, and establishing the ubiquity of sampling problems across a broad range of subject areas. We note that a quantum computer can be used as a sample source and thereby gain insights into quantum-mechanical processes and distributions.
After introducing a sampling algorithm as a transformation of uniformly distributed random bits into bits distributed according to a non-uniform probability distribution, we then briefly compare the implementation of random number generation and sampling algorithms in the classical versus the quantum realm, establishing that quantum computers need a different approach due to their larger internal state space.

Subsequently, we describe how the previously covered properties of quantum computing can be utilized to improve some well-known solutions approaches to sampling problems.
The first of these solution approaches treats the problem of Gaussian Boson Sampling, in which a coherent superposition of input states is to be transformed into a potentially high-dimensional superposition following a multi-modal probability distribution, from which a fair sample is to be taken. While the problem itself has not definitively been assigned a complexity class, current protocols solving it using a photonic quantum computer have been proven to be in the “hard” complexity class $\#P$ and can be used to perform a probability chain rule method or Metropolis independence sampling within minutes, while their classical simulation takes months.

In the second approach we explain quantum-enhanced Markov Chain Monte Carlo methods, specifically a Metropolis-Hastings algorithm, which can be used, e.g., for combinatorial optimization or to sample from probability distributions that are classically hard to access such as the Boltzmann distribution. The quantum-enhanced version of this algorithm uses time-evolution for the step of proposing candidates for a state update, thereby achieving faster convergence than the classical version of the algorithm. Whether the tuning of the free parameters could further improve the performance and how other quantum circuits would perform in the place of the time-evolution remains an open research question.

The third approach considers Variational Monte Carlo structures. Here, parametrized wave function ansätze are used, either fitting the physical problem to be solved or represented by a neural network. Observables and quantities can subsequently be computed by sampling from the probability distribution of the wave function, i.e., the absolute square of the wave function factors. VMC has the drawback that the ansätze need to be computed both classically and on the quantum devices. To date, only a single ansatz has been used. Here, of course, research for new ansätze can be pursued. Yet it is still unclear whether VMC are preferable to similar approaches such as VQE.

Similar to the previous approach, we cover Boltzmann Machines as wave function ansätze and as generative machine learning models. BMs need sampling from both a model and a data Boltzmann distribution in order to optimize their parameters to make the former match the latter as closely as possible using gradient descent. Here, quantum computers are employed to model the Boltzmann distribution such that sampling naturally becomes measuring in the $z$-basis.
We cover several methods in BM wave function construction, both using circuit-based constructions and quantum annealing.
A handful of indications that faster training and better approximation can be achieved have been given.
Yet, circuit construction is not considered NISQ-friendly, despite having quadratic complexity in space and time requirements.
Furthermore, augmenting the Boltzmann Distribution by including a transverse field Hamiltonian to the Ising Energy has been discussed. Numerical studies with perfect circuit construction demonstrate higher expressibility, but a general advantage has not been proven.

Lastly, we describe Bayesian networks, being directed acyclic graphs that describe chained sets of probabilities, where the probability of a child node depends directly on that of his parents. These networks can be directly mapped to quantum circuits, the probabilities being represented by the qubits and the edges by the control wires between them that steer unitaries acting on them. The usage of these quantum Bayesian networks might provide additional insights into the structure of relationships be-
tween different influence factors and may be suitable to accommodate the higher expressive power of dynamic Bayesian networks using higher dimensional space of quantum computers.

To summarize, what most of these approaches have in common is that they use the inherent randomness of a quantum computer expressed by its ability to store basis states in a superposition that collapses to each of them with a certain probability when measured, as well as the high dimensionality of its state spaces, to model complicated, possibly classically intractable probability distributions, and subsequently sample from them. 

However, for most of these approaches, the hardness in terms of theoretical complexity is unknown, as well as whether certain quantum approaches might be working better for them than others, and, if so, why. In a lot of cases, it is not even clear why exactly the presented quantum approach outperforms the comparable classical approach  from a theoretical perspective. Furthermore, many details about how to practically implement the approaches on near-term quantum computers in an optimal, efficient manner, i.a. the optimal way to tune quantum-specific hyperparameters, are not entirely clear yet. By presenting these open research questions, this paper lays the groundwork for further exploration regarding the applicability of quantum computing in sampling applications, as we and our readers, now more aware of them, can strive to answer them in our future work.

\section*{Acknowledgements}

The authors acknowledge funding from the German Federal Ministry of Education and Research under the funding program "Förderprogramm Quantentechnologien – von den Grundlagen zum Markt" (funding program quantum technologies – from basic research to market), project BAIQO, 13N16089 (Maximilian Balthasar Mansky) and project Q-Grid, 13N16177 (David Bucher), as well as from the German Federal Ministry for Economic Affairs and Climate Action, project PlanQK, 01MK20005I (Daniëlle Schuman).



\bibliographystyle{IEEEtran}
\bibliography{locallib}

\end{document}

%% file: acronyms.tex
\newacro{MCMC}{Markov Chain Monte Carlo}
\newacro{DMRG}{Density Matrix Renormalization Group}
\newacro{GBS}{Gaussian Boson Sampling}
\newacro{VMC}{Variational Monte Carlo}
\newacro{RBM}{Restricted Boltzmann Machine}
\newacro{NNQS}{Neural Network Quantum States}

%% file: monte_carlo.tex
\subsection{Quantum-Enhanced Markov Chain Monte Carlo}\label{sec:mc-enhanced}

Markov Chain Monte Carlo (MCMC) is a statistical approach for generating random samples from a target probability distribution.
The basic idea of Markov Chains is to start from an initial state and repeatedly jump to new states according to a transition rule.
This allows for a computationally inexpensive estimation of various statistics (e.g., mean, variance) of the target distribution.
In various parts of physics, they are widely used to estimate observables of statistical systems whose probability distributions are inaccessible through direct computation~\cite{Huang1987-xd}. Furthermore, they are used for sampling from Boltzmann distributions~\cite{ackley1985} (see also Sec.~\ref{sec:boltzmann}) and for combinatorial optimization using the simulated annealing heuristic~\cite{kirkpatrick1983}.

Sampling from the Boltzmann distribution of a classical Ising spin-glass at low temperatures is known to be a hard problem~\cite{barahona1982} The probability of a certain classical spin string $\mathbf{s} = \{\pm 1\}^N$ is given by
\begin{align}
    p(\mathbf{s}) = \frac{1}{\mathcal{Z}} e^{-\beta E(\mathbf{s})},
\end{align}
where $\beta$ is the reciprocal temperature given by $\beta^{-1} = k_B T$, with $k_B$ being the Boltzmann factor. The energy of the general Ising system~\cite{ising1925} is given by
\begin{align}\label{eq:ising_spin_glass}
    E(\mathbf{s}) = - \sum_{i<j} J_{ij} s_i s_j - \sum h_i s_i.
\end{align}
The partition function $\mathcal{Z}$ is defined as the sum over the Boltzmann factors $\mathcal{Z} = \sum_{\{\mathbf{s}\}} e^{-\beta E(\mathbf{s})}$. Although the Boltzmann factors are easy to compute independently, the partition function is not, because of the exponential number of summands. Without specific domain knowledge and advanced analytical methods, the calculation is intractable.

The Metropolis-Hastings algorithm is an MCMC method to sample from many distributions, like this Boltzmann distribution. Essentially, we start at a random spin configuration and then propose update steps. These are accepted or rejected according to the transition probabilities~\cite{metropolis1953, hastings1970}. The acceptance probability of a proposed update $\mathbf{s} \rightarrow \mathbf{s}'$ is given by
\begin{align}\label{eq:acceptance}
    A(\mathbf{s}, \mathbf{s}') = \min \left\{1, \frac{p(\mathbf{s}') Q(\mathbf{s}|\mathbf{s}')}{p(\mathbf{s}) Q(\mathbf{s}'|\mathbf{s})}\right\},
\end{align}
with $Q(\mathbf{s}|\mathbf{s}')$ being the transition probability of the system moving from $\mathbf{s}$ to $\mathbf{s}'$. These transition probabilities are chosen such that the samples from this procedure resemble the demanded probability distribution.

Commonly, the transition probabilities are chosen symmetrically, meaning that the probability is the same no matter if the state is moving from $\mathbf{s} \rightarrow \mathbf{s}'$ or the other way round. In the acceptance probability \eqref{eq:acceptance} the expression then reduces to just a fraction of the state probabilities. Furthermore, note that due to that ratio, we never need to calculate the partition function explicitly. A metric to measure how well the Markov chain samples through the probability distribution is the acceptance ratio $\alpha$, i.e., the number of accepted proposals compared to the number of all trials.
Theoretical investigations suggest an optimal value of $\alpha = 0.243$ for random walk problems~\cite{gelman1997}. However, most importantly, the acceptance ratio should not drop close to zero in order to maintain a good sampling quality. Small acceptance rates are an indication that the chain gets stuck in a trench in the energy landscape.

\begin{figure}
    \centering
    \input{energie-drawing}
    \caption{Visualization of different candidate proposal techniques. The local one does achieve relatively good acceptance rates while not exploring the state space. Uniform updating tries to explore the state space but struggles with acceptance since the proposed state has most likely high energy. However, the discussed quantum proposal routine samples states that are far away in the state space while having comparable energy, thus, also leading to high acceptance rates.} 
    \label{fig:mcmc_enhancement}
\end{figure}

Typically, update-proposal strategies comprise
\begin{itemize}
    \item \emph{Local Spin flips}: 
    Here, just a random spin is chosen to be flipped. In many scenarios, mainly where the energy landscape is very rugged, single spin flips cannot get the chain out of deep trenches, also depicted in Fig.~\ref{fig:mcmc_enhancement}. Since only a limited number of proposals are available, if all of them are unlikely to be accepted due to higher energy, the chain gets stuck, and the acceptance rate drops.
    \item \emph{Uniform updates}: Now, update candidates $\mathbf{s}'$ are chosen randomly. In higher temperature simulations, this strategy works fine and is able to traverse the whole state space relatively quickly, which means fast convergence of the Markov chain. However, the acceptance rate drops rapidly as the temperature decreases~\cite{layden2022}, see Fig.~\ref{fig:mcmc_enhancement}.
    \item \emph{Cluster updates}: In the phase transition between the magnetized and disordered state of the Ising model, generally, ordered patches emerge in the material. For this reason, cluster update proposals have been introduced~\cite{wolff1989, houdayer2001}. Being able to explore the state space rather quickly with high acceptance rates, they only work in critical phases of the material and similarly lose their advantage as soon as the temperature falls significantly below the critical point.
\end{itemize}
In general, sampling from low-temperature Ising spin glasses with Markov chains is plagued by slow convergence and, therefore, long runtimes.

To counteract this problem, Layden et al.~\cite{layden2022} suggested a quantum routine to propose updates. The quantum register is prepared in the state of the current MCMC chain and undergoes an arbitrarily chosen unitary evolution. The authors of~\cite{layden2022} chose to use a time evolution of the problem Hamiltonian paired with a mixer Hamiltonian known from the Quantum Approximate Operator Ansatz (QAOA)~\cite{farhi2014a, hadfield2019}. The joint Hamiltonian can be expressed through \begin{align} H_\text{int} = (1 - \gamma) \alpha \sum_{\mathbf{s}} E(\mathbf{s})\ket{\mathbf{s}}\bra{\mathbf{s}} - \gamma \sum_i \sigma^x_i,
\end{align}
where $\alpha$ is a normalizing factor and $\gamma \in [0, 1]$ controls the strength of the quantum transitions by increasing the effect of the mixer.

The output of measuring the state
\begin{align}
    \exp(-i H_{\text{int}} t) \ket{\mathbf{s}} \approx \prod^T \exp(-i H_\text{int} t / T)\ket{\mathbf{s}}
\end{align}
is the proposed spin configuration, where the Hamiltonian is time-evolved using Trotterization.
Important to note is the symmetry of the transition probability $|\bra{\mathbf{s}} U \ket{\mathbf{s}'}| = |\bra{\mathbf{s}'} U \ket{\mathbf{s}}|$ using this approach.
Yet, there are still two free parameters $\gamma$ and $t$ that need to be set. To circumvent the need for tuning them, \cite{layden2022} chose to sample the parameters of each Monte Carlo iteration randomly, decreasing the bias of a constant setting. The achieved effect, compared to local and uniform updating, is visualized in Fig.~\ref{fig:mcmc_enhancement}.

The authors of~\cite{layden2022} have found a significant improvement in convergence speed when compared to local and uniform updating procedures. Furthermore, with clever error mitigation in use, they have also been able to observe a performance increase when running on quantum hardware. The gain was not as big as the simulations suggested, nevertheless, faster convergence than local and uniform updates has been achieved.

Remarkably, the proposed algorithm never miscalculates a quantity based on quantum imperfections since the precise values are computed using the Metropolis algorithm. The quantum routine only produces  update proposals, which still need to be accepted in order to be included to the computation. The quantum routine only helps with increasing the acceptance rate and the exploration speed, leading to faster convergence.

How the authors~\cite{layden2022} deal with the parameters in the time evolution surely removes bias but is not the ideal setting. Thus more effort can be devoted to parameter tuning. Furthermore, performing a time evolution on a sample is heuristically chosen and has no physical reason. Thus other quantum proposal finding methods can be investigated, like a reverse quantum annealing ansatz or a Quantum Phase Estimation (QPE)-based method, as already mentioned in~\cite{layden2022}.

\subsection{Variational Monte Carlo}

One of the most demanding challenges in modern physics, chemistry, and material sciences is the description and simulation of many-body quantum systems and their ground state at zero or low temperatures~\cite{carleo2017}.
Tensor network methods, like density matrix renormalization group (DMRG)~\cite{schollwoeck_2011}, projected entangled pair states (PEPS)~\cite{orus2014} or multi-scale renormalization ansatz (MERA)~\cite{jia2019}, have proven incredibly valuable for computing observables for many classes of materials and chemical systems. Despite their success, especially in one-dimensional and, with limitations, in two-dimensional materials~\cite{eisert2010}, tensor networks reach practical computational boundaries in higher-dimensional lattices and highly entangled quantum states~\cite{jia2019}. 

On the other hand, Quantum Monte Carlo Methods (QMC) also emerged as a direct calculation of quantum mechanical properties through approximation with Monte Carlo sampling. Due to the infamous sign problem, they struggle to simulate fermionic and frustrated systems~\cite{loh1990, li2019, troyer2005, ceperley1977}, such as high-temperature superconductivity~\cite{li2019} or electronic band structure computations~\cite{xia2018}.
Both are extremely interesting systems, not only for theoretical investigation but also for practical use.
Recently, an QC-assisted QMC method has been proposed to control the effect of sign problem~\cite{huggins2022}. Notably, it achieves comparable results to state-of-the-art classical methods.

Variational Monte Carlo (VMC)~\cite{mcmillan1965,carleo2017} methods evade the sign problem by using parametrized wave function ansätze. In comparison to the previously discussed unbiased ansätze, VMC is inherently biased~\cite{jia2019}, as the wave function ansatz has to be meticulously chosen to fit the problem one wants to solve. The parameters of the quantum wave function ansatz are subsequently optimized using an MC evaluation of the energy of the system.

Given a local Hamiltonian $H$, i.e., a sum of Hamiltonians $H_a$ who act only on a subspace of complete system Hilbert space $\mathcal{H}_a \subset \mathcal{H}$ with $\mathrm{dim}\,\mathcal{H}_a \in \mathcal{O}(1)$. We typically consider nearest-neighbor Hamiltonians, whose local Hamiltonians act on two spins or qubits.

Let $\ket{\psi_\Omega} = \sum_{\{\mathbf{x}\}} \psi_\Omega(\mathbf{x}) \ket{\mathbf{x}}$ be the wave function ansatz, with the parameters gathered in $\Omega$ and $\mathbf{x} \in \{-1, +1\}^N$. We can evaluate the expectation value of a local observable or Hamiltonian efficiently using MC sampling:
\begin{align}
    E(\Omega) &= \bra{\psi_\Omega} H \ket{\psi_\Omega} \\ &= \sum_a \sum_{\{\mathbf{x}\}, \{\mathbf{x}'\}} \psi_\Omega^*(\mathbf{x}) \psi_\Omega(\mathbf{x}') \bra{\mathbf{x}}H_a\ket{\mathbf{x}'} \\
    &= \sum_{\{\mathbf{x}\}} |\psi_\Omega(\mathbf{x})|^2 E^\text{loc}_\Omega(\mathbf{x}) \approx 
    \frac{1}{N_{MC}} \sum_{i}^{N_{MC}} E^\text{loc}_\Omega(\mathbf{x}_i),
\end{align}
where $\mathbf{x}_i$ is sampled from the probability distribution $|\psi_\Omega(\mathbf{x})|^2$. This is only possible if $E^\text{loc}_\Omega(\mathbf{x})$ is efficiently computable on classical computers,
\begin{align}
    E^{\text{loc}} = \sum_a \sum_{\mathbf{x}'} \frac{\psi_\Omega(\mathbf{x}')}{\psi_\Omega(\mathbf{x})} \bra{\mathbf{x}} H_a \ket{\mathbf{x}'} = \sum_a \frac{\bra{\mathbf{x}} H_a \ket{\psi_\Omega} }{\psi_\Omega(\mathbf{x})}.
\end{align}
Typically, this is the case for a sufficiently well-behaved ansatz $\psi_\Omega$~\cite{carleo2017}.

Usually, the samples are drawn from an MCMC simulation, but this simulation is prone to similar issues as previously discussed in Sec.~\ref{sec:mc-enhanced}. However, as the probability distribution is of an entangled quantum state and not of a classical Ising spin-glass, we cannot utilize the previously presented method for enhancing the MCMC. Instead, we propose to prepare the quantum state directly on the QC and measure in the $z$-basis, similar to~\cite{xia2018}. This automatically produces samples that are distributed according to $|\psi_\Omega(\mathbf{x})|^2$.
However, for this approach to be viable, we need a circuit $U_\Omega$ comprised of a polynomial number of gates, such that $U_\Omega \ket{0} = \ket{\psi_\Omega}$.

Furthermore, in order to train the parameters of the wave function ansatz, we need an efficient method of computing the gradient with respect to the parameters.
The gradient must be representable as a set of local Hermitian operators $O_\lambda$, where
\begin{align}
 O_\lambda(\mathbf{x}) = \frac{\partial}{\partial \Omega_\lambda} \log \psi_\Omega(\mathbf{x})
\end{align}
From that follows the energy gradient instantly, following~\cite{sorella2007}:
\begin{align}
\frac{\partial E(\Omega) }{\partial \Omega_\lambda} = \langle O_\lambda^\dagger H \rangle - \langle O_\lambda^\dagger \rangle \langle H \rangle.
\end{align}
Finally, gradient-descent-based optimization can be pursued after the gradient has been approximated with MC using the same samples that already have been obtained for computing the energy~\cite{sorella2007}.

Typically, physically motivated wave function ansätze are used~\cite{mcmillan1965, sorella1998, sorella2007}, but recently efforts have been made to unbias the ansatz through universal function approximators, also known as neural networks~\cite{carleo2017, choo2019, ferrari2019}. As neural networks are well known for various other applications, they are a fitting choice, especially since classical computational acceleration is possible.

At this point, it makes sense to draw a dividing line to the well-known Variational Quantum Eigensolver (VQE)~\cite{mcclean2016, tilly2022}, which is likewise used to find the ground state of quantum many-body systems. Both methods rely on parametrized circuits, which aim to encode a physical quantum state. VQE computes the energy directly on the Quantum Device and has to perform multiple measurements to obtain the gradient with respect to the parameters~\cite{tilly2022}. On the other hand, the VMC ansatz is only used for sampling, energy and gradients can are computed classically. Furthermore, one can use higher order methods, like Stochastic Reconfiguration~\cite{sorella1998} for faster training, which are otherwise only possible with far more circuit evaluations~\cite{stokes2020}. However, these benefits come with the drawback that the ansatz wave functions are required to be efficiently prepared on the QC and simultaneously be efficiently evaluated on classical computers. Restricted Boltzmann Machine (RBM) based ansätze could be a fitting candidate, as they have been used for VMC~\cite{carleo2017} and polynomial size RBM wave function construction circuits exist~\cite{xia2018}. 

Finally, it is an open question whether it is beneficial to restrict variational QC to states that are efficiently produced with classical computers in order to make use of the mentioned advantages. Furthermore  –– except for the RBM quantum state further discussed in the next section ––  the authors of this paper are not aware of other suitable ansätze.

\subsection{Quantum Boltzmann Machines}\label{sec:boltzmann}

Boltzmann Machines (BM) are generative machine learning models that have been introduced by Hinton et al.~\cite{hinton1986}. 
By turning the interactions in a classical Ising spin-glass into parameters $\{J, h\}$, the Boltzmann distribution of the parameterized Ising energy becomes a learnable probability distribution over a discrete domain~\cite{hinton2012}. The spin variables $s_i \in \{-1, 1\}$ are the discrete inputs (and outputs) to the network, see the left-hand side of Fig.~\ref{fig:boltzmann}, while the learnable probability distribution has the formula:
\begin{align}
    p(\mathbf{x}) = \frac{1}{\mathcal{Z}}  e^{-\beta E_{J, h}(\mathbf{x})}.
\end{align}
Here, the inverse temperature $\beta$ serves as a regularization parameter, governing the ruggedness of the resulting probability distribution. $\mathcal{Z}$ and $E_{J, h}(\mathbf{x})$ are defined as described in Sec. \ref{sec:mc-enhanced}.
To learn the probability distribution of a set of data points, one can tune the parameters using gradient descent. 
This makes them applicable in a variety of Machine Learning (ML) tasks, including unsupervised learning~\cite{hinton2012, hinton2006a}, supervised learning~\cite{larochelle2012}
and even reinforcement learning problems~\cite{crawford2019}. As such, application areas range from %
image recognition and denoising~\cite{tang2012} to 
cybersecurity~\cite{sarker2021} and medical diagnostics~\cite{vantulder2014}.

\begin{figure}
    \centering
    \begin{tikzpicture}[node/.style={fill=orange!10,draw=black,thick,circle, inner sep=0pt, minimum width=0.5cm}, weights/.style={draw=black!70}]
        \foreach \i in {0,...,9} {
            \foreach \j in {\i,...,8} {
                \draw[weights] (\i * 40 + 17:1.3cm) -- (\j * 40 + 57:1.3cm); 
            }
        }
        \foreach \i in {0,...,9} {
            \node[node] at (\i * 40 +17:1.3cm) {\small $x_{\i}$};
        }
    \end{tikzpicture}
    \hspace{10pt}
    \begin{tikzpicture}[node/.style={fill=orange!10,draw=black,thick,circle, inner sep=0pt, minimum width=0.5cm}, weights/.style={draw=black!70}]
        \foreach \i in {0,...,4} {
            \foreach \j in {0,...,5} {
                \draw[weights] (\i * 0.7cm + 0.35cm, 1.2) -- (\j * 0.7cm, -1.2); 
            }
        }
        \foreach \i in {0,...,5} {
            \node[node] at (\i * 0.7cm, -1.2) {\small $v_{\i}$};
        }
        \foreach \i in {0,...,4} {
            \node[node,fill=blue!10] at (\i * 0.7cm + 0.35cm, 1.2) {\small $h_{\i}$};
        }
    \end{tikzpicture}
    \caption{Drawing of the Boltzmann Machine (left) and Restricted Boltzmann Machine (right). Connections between the nodes indicate the Boltzmann weights $J_{ij}$, each node itself contains a bias weight $h_i$. In the RBM, the nodes are split up into two groups: Hidden and visible units. It has no connections between two visible (hidden) units.}
    \label{fig:boltzmann}
\end{figure}

For training, the Kullback-Leibler-Divergence between the BM's distribution and the dataset's distribution is chosen as the loss function. In order to obtain the gradients with respect to the parameters of the Boltzmann Machine, one needs sample from both these probability distributions~\cite{hinton2012}.
Done classically, exact sampling from these distributions can become exceedingly time-consuming since one has to compute the partition function. Similar to Sec.~\ref{sec:mc-enhanced}, Metropolis sampling can be utilized to alleviate this issue. In fact, the BM problem is exactly the same as the one previously discussed. Although proposed in~\cite{layden2022}, no experiments have so far been conducted on applying the quantum-enhanced method described in Sec.~\ref{sec:mc-enhanced} to sampling from a BM.

To achieve better classical performance, one typically divides the in- and output units of the BM into two separate clusters, called visible and hidden units~\cite{hinton2006a}, denoted with $\mathbf{v}$ and $\mathbf{h}$, which can be seen in the right-hand drawing of Fig.~\ref{fig:boltzmann}. When disallowing any connections between units of the same cluster, the emerging Restricted Boltzmann Machine (RBM) gains some advantages. First, it is easier to sample using Gibbs sampling, another MCMC sampling method that updates hidden and visible units in an alternating fashion~\cite{geman1984, hinton2012}. Furthermore, training using Contrastive Divergence learning becomes way more efficient~\cite{pmlr-vR5-carreira-perpinan05a}. As the naming suggests, the visible units are the ones exposed to the outside, i.e., they need to have the same dimensionality as the input data. Hidden units, on the other hand, are traced over and only referenced internally. The probability distribution over visible units is defined as~\cite{hinton2012}
\begin{align}
    p(\mathbf{v}) = \frac{1}{\mathcal{Z}} \sum_{\{\mathbf{h}\}} e^{-\beta E_{J,h}(\{\mathbf{v}, \mathbf{h}\})}
\end{align}
where $J$ incorporates the required connectivity restrictions. Despite the initial drawback of reduced expressive power due to the limitations in connectivity, one can still adapt the RBM's expressiveness by choosing an appropriate number of hidden units~\cite{montufar2014}. 
Nevertheless, hidden units can likewise be added to fully connected BMs.

Quantum Computing applications for Boltzmann machines are twofold: first, encoding the Boltzmann probability in quantum states, and second, creating a modified BM, called Quantum Boltzmann Machine (QBM), that adds a transverse field to the BM's Ising energy function.

In the first approach, we aim to prepare a quantum state $\ket{\psi} = \sum_{\{\mathbf{x}\}} \psi(\mathbf{x}) \ket{\mathbf{x}}$ on the computer. Wiebe et al.~\cite{wiebe2015} assembled a circuit for preparing such states for RBMs that is based on a mean-field approximation with followed importance sampling. They observed a quadratic reduction of training data access during the optimization process.
Further state preparation methods include an approximate imaginary time evolution method based ansatz called VarQITE~\cite{zoufal2021}, and Quantum Annealing (QA)-based approaches~\cite{amin2015, benedetti2016}. The QA approaches need precise tuning of the inverse temperature parameter~\cite{adachi2015}. Otherwise, the final distribution will not be close to a Boltzmann distribution~\cite{amin2015}. 

Also belonging to the state preparation methods, but with the alternative purpose of finding the ground state of specific Hamiltonians are the VMC methods, where we can now close the circle to Sec.~\ref{sec:monte-carlo}. Here, Xia et al.~\cite{xia2018} proposed an RBM quantum state circuit for electronic structure calculations. Instead of considering just amplitude information, they, and related methods~\cite{sajjan2021, sureshbabu2021}, additionally include sign or phase information through a separate network, i.e., $\ket{\psi} = \sum_{\{\mathbf{x}\}} s(\mathbf{x}) \sqrt{p(\mathbf{x})} \ket{\mathbf{x}}$. The circuit construction works similarly to~\cite{wiebe2015}, however, no mean-field initialization is present. Yet, simulations of the RBM approach produce good results on small molecules, like $\text{H}_2$, $\text{Li}\text{H}$ or $\text{H}_2\text{O}$, outperforming the Hartree-Fock outcome.
Despite having quadratic space and depth requirements, the NISQ applicability is criticized because of a huge constant overhead in circuit construction and the required ability to perform in-circuit measurements, which is not necessarily given in NISQ devices~\cite{paul2023}. Furthermore, the expressive power of the sign part is also questioned, but this is also a known issue in VMC literature~\cite{bukov2021, park2022, westerhout2020}, even with different ansätze for the phase information.

The second approach augments the original Ising Hamiltonian with an additional transverse field into the spin-glass Hamiltonian. The Quantum Boltzmann machine leverages quantum effects to encode distributions that are believed to be more expressive than ordinary BMs~\cite{amin2018}
\begin{align}
    H(J, h, \Gamma) = -\sum_{ij} J_{ij} \sigma^z_i \sigma^z_j - \sum_i h_i \sigma^z_i - \sum_i \Gamma_i \sigma^x_i.
\end{align}
Here, $\sigma_i^{x}, \sigma_i^{z}$ denote the Pauli $x$- and $z$-operators at qubit $i$. In numerical studies, assuming perfect state preparation, QBMs seem to outperform classical BMs~\cite{amin2018}. Nonetheless, one major advantage of BMs, that is direct access to the analytical gradient, is unfortunately lost in QBMs due to the non-commuting terms in the Hamiltonian. 

In conclusion, BMs on quantum computers pave interesting directions. Advanced circuit generation for machine learning, as well as VMC application, needs further investigation as current methods are not satisfactory with regards to practical applicability~\cite{paul2023}. For QBMs, it is the question if they can withstand normal BMs in practical circumstances. A general performance benefit has to our knowledge so far only been demonstrated on a few small instances~\cite{amin2018, Benedetti17}.

%% file: energie-drawing.tex
\begin{tikzpicture}

\draw[<->] (0,3)  -- node[left]{$E(\mathbf{s})$} (0,0) -- node[below] {$\mathbf{s}$} (7,0) ;
\draw plot[smooth] coordinates {(0,1.5) (.2, 3) (.4, .8) (.5, 1.2) (1, 0.6) (1.5, 1.9) (1.7, 1.7) (2, 2.2) (2.2, 2.1) (2.6, 2.8) (2.9, 2.2) (3.5, 1.8) (4,2.4) (4.5, 1.7) (5, 2.5) (5.5, .6) (6, 1.7) (6.2, 1.6) (6.4, 1.8) (6.7, .4) (6.9, 1.4) (7,1.5)};
\fill (1, .6) circle (.05) node (local) {};
\node at (local) [above = .5cm, blue] {Local};
\draw[->, shorten > = 5pt, blue, thick] (local) -- (.5, 1.2);
\draw[->, shorten > = 5pt, gray, thick] (local) -- (4.5, 1.7) node[midway, below, sloped] {Uniform};
\draw[->, shorten > = 5pt, red, thick] (local) to[bend right=10] (5.5, .6);
\draw(3.2, .15) node[red] {Quantum};

\end{tikzpicture}

%% file: main.bbl
\begin{thebibliography}{10}
\providecommand{\url}[1]{#1}
\csname url@samestyle\endcsname
\providecommand{\newblock}{\relax}
\providecommand{\bibinfo}[2]{#2}
\providecommand{\BIBentrySTDinterwordspacing}{\spaceskip=0pt\relax}
\providecommand{\BIBentryALTinterwordstretchfactor}{4}
\providecommand{\BIBentryALTinterwordspacing}{\spaceskip=\fontdimen2\font plus
\BIBentryALTinterwordstretchfactor\fontdimen3\font minus
  \fontdimen4\font\relax}
\providecommand{\BIBforeignlanguage}[2]{{%
\expandafter\ifx\csname l@#1\endcsname\relax
\typeout{** WARNING: IEEEtran.bst: No hyphenation pattern has been}%
\typeout{** loaded for the language `#1'. Using the pattern for}%
\typeout{** the default language instead.}%
\else
\language=\csname l@#1\endcsname
\fi
#2}}
\providecommand{\BIBdecl}{\relax}
\BIBdecl

\bibitem{witte_statistics_2017}
R.~S. Witte and J.~S. Witte, \emph{Statistics}.\hskip 1em plus 0.5em minus
  0.4em\relax John Wiley \& Sons, 2017.

\bibitem{agresti_art_2007}
A.~Agresti and C.~Franklin, ``The art and science of learning from data,''
  \emph{Upper Saddle River, New Jersey}, vol.~88, 2007.

\bibitem{thompson_sampling_2012}
S.~K. Thompson, \emph{Sampling}.\hskip 1em plus 0.5em minus 0.4em\relax John
  Wiley \& Sons, 2012, vol. 755.

\bibitem{nielsen_quantum_2010}
M.~A. Nielsen and I.~L. Chuang, \emph{\BIBforeignlanguage{en}{Quantum
  computation and quantum information}}, 10th~ed.\hskip 1em plus 0.5em minus
  0.4em\relax Cambridge ; New York: Cambridge University Press, 2010.

\bibitem{madsen_quantum_2022}
\BIBentryALTinterwordspacing
L.~S. Madsen, F.~Laudenbach, M.~F. Askarani, F.~Rortais, T.~Vincent, J.~F.~F.
  Bulmer, F.~M. Miatto, L.~Neuhaus, L.~G. Helt, M.~J. Collins, A.~E. Lita,
  T.~Gerrits, S.~W. Nam, V.~D. Vaidya, M.~Menotti, I.~Dhand, Z.~Vernon,
  N.~Quesada, and J.~Lavoie, ``Quantum computational advantage with a
  programmable photonic processor,'' \emph{Nature}, vol. 606, no. 7912, pp.
  75--81, Jun. 2022, number: 7912 Publisher: Nature Publishing Group. [Online].
  Available: \url{https://www.nature.com/articles/s41586-022-04725-x}
\BIBentrySTDinterwordspacing

\bibitem{lund_quantum_2017}
\BIBentryALTinterwordspacing
A.~P. Lund, M.~J. Bremner, and T.~C. Ralph, ``Quantum sampling problems,
  {BosonSampling} and quantum supremacy,'' \emph{npj Quantum Information},
  vol.~3, no.~1, pp. 1--8, Apr. 2017, number: 1 Publisher: Nature Publishing
  Group. [Online]. Available:
  \url{https://www.nature.com/articles/s41534-017-0018-2}
\BIBentrySTDinterwordspacing

\bibitem{haferkamp_linear_2022}
\BIBentryALTinterwordspacing
J.~Haferkamp, P.~Faist, N.~B.~T. Kothakonda, J.~Eisert, and N.~Yunger~Halpern,
  ``Linear growth of quantum circuit complexity,'' \emph{Nature Physics},
  vol.~18, no.~5, pp. 528--532, May 2022. [Online]. Available:
  \url{https://www.nature.com/articles/s41567-022-01539-6}
\BIBentrySTDinterwordspacing

\bibitem{herrero-collantes_quantum_2017}
\BIBentryALTinterwordspacing
M.~Herrero-Collantes and J.~C. Garcia-Escartin, ``Quantum random number
  generators,'' \emph{Reviews of Modern Physics}, vol.~89, no.~1, p. 015004,
  Feb. 2017, publisher: American Physical Society. [Online]. Available:
  \url{https://link.aps.org/doi/10.1103/RevModPhys.89.015004}
\BIBentrySTDinterwordspacing

\bibitem{AA}
\BIBentryALTinterwordspacing
S.~Aaronson and A.~Arkhipov, ``The computational complexity of linear optics,''
  ser. STOC '11.\hskip 1em plus 0.5em minus 0.4em\relax New York, NY, USA:
  Association for Computing Machinery, 2011, p. 333–342. [Online]. Available:
  \url{https://doi.org/10.1145/1993636.1993682}
\BIBentrySTDinterwordspacing

\bibitem{Zhong}
\BIBentryALTinterwordspacing
H.-S. Zhong, L.-C. Peng, Y.~Li, Y.~Hu, W.~Li, J.~Qin, D.~Wu, W.~Zhang, H.~Li,
  L.~Zhang, Z.~Wang, L.~You, X.~Jiang, L.~Li, N.-L. Liu, J.~P. Dowling, C.-Y.
  Lu, and J.-W. Pan, ``Experimental gaussian boson sampling,'' \emph{Science
  Bulletin}, vol.~64, no.~8, pp. 511--515, apr 2019. [Online]. Available:
  \url{https://doi.org/10.1016\%2Fj.scib.2019.04.007}
\BIBentrySTDinterwordspacing

\bibitem{Lund}
\BIBentryALTinterwordspacing
A.~Lund, A.~Laing, S.~Rahimi-Keshari, T.~Rudolph, J.~O'Brien, and T.~Ralph,
  ``Boson sampling from a gaussian state,'' \emph{Physical Review Letters},
  vol. 113, no.~10, sep 2014. [Online]. Available:
  \url{https://doi.org/10.1103\%2Fphysrevlett.113.100502}
\BIBentrySTDinterwordspacing

\bibitem{Hamilton}
\BIBentryALTinterwordspacing
C.~S. Hamilton, R.~Kruse, L.~Sansoni, S.~Barkhofen, C.~Silberhorn, and I.~Jex,
  ``Gaussian boson sampling,'' \emph{Physical Review Letters}, vol. 119,
  no.~17, oct 2017. [Online]. Available:
  \url{https://doi.org/10.1103\%2Fphysrevlett.119.170501}
\BIBentrySTDinterwordspacing

\bibitem{Kruse}
\BIBentryALTinterwordspacing
R.~Kruse, C.~S. Hamilton, L.~Sansoni, S.~Barkhofen, C.~Silberhorn, and I.~Jex,
  ``Detailed study of gaussian boson sampling,'' \emph{Physical Review A}, vol.
  100, no.~3, sep 2019. [Online]. Available:
  \url{https://doi.org/10.1103\%2Fphysreva.100.032326}
\BIBentrySTDinterwordspacing

\bibitem{Zhong20}
\BIBentryALTinterwordspacing
H.-S. Zhong, H.~Wang, Y.-H. Deng, M.-C. Chen, L.-C. Peng, Y.-H. Luo, J.~Qin,
  D.~Wu, X.~Ding, Y.~Hu, P.~Hu, X.-Y. Yang, W.-J. Zhang, H.~Li, Y.~Li,
  X.~Jiang, L.~Gan, G.~Yang, L.~You, Z.~Wang, L.~Li, N.-L. Liu, C.-Y. Lu, and
  J.-W. Pan, ``Quantum computational advantage using photons,'' \emph{Science},
  vol. 370, no. 6523, pp. 1460--1463, 2020. [Online]. Available:
  \url{https://www.science.org/doi/abs/10.1126/science.abe8770}
\BIBentrySTDinterwordspacing

\bibitem{Bulmer}
\BIBentryALTinterwordspacing
J.~F.~F. Bulmer, B.~A. Bell, R.~S. Chadwick, A.~E. Jones, D.~Moise, A.~Rigazzi,
  J.~Thorbecke, U.-U. Haus, T.~V. Vaerenbergh, R.~B. Patel, I.~A. Walmsley, and
  A.~Laing, ``The boundary for quantum advantage in gaussian boson sampling,''
  \emph{Science Advances}, vol.~8, no.~4, p. eabl9236, 2022. [Online].
  Available: \url{https://www.science.org/doi/abs/10.1126/sciadv.abl9236}
\BIBentrySTDinterwordspacing

\bibitem{Huang1987-xd}
K.~Huang, \emph{Statistical Mechanics}, 2nd~ed.\hskip 1em plus 0.5em minus
  0.4em\relax {John Wiley \& Sons}, 1987.

\bibitem{ackley1985}
\BIBentryALTinterwordspacing
D.~H. Ackley, G.~E. Hinton, and T.~J. Sejnowski, ``\BIBforeignlanguage{en}{A
  learning algorithm for boltzmann machines},''
  \emph{\BIBforeignlanguage{en}{Cognitive Science}}, vol.~9, no.~1, pp.
  147--169, Jan. 1985. [Online]. Available:
  \url{https://www.sciencedirect.com/science/article/pii/S0364021385800124}
\BIBentrySTDinterwordspacing

\bibitem{kirkpatrick1983}
\BIBentryALTinterwordspacing
S.~Kirkpatrick, C.~D. Gelatt, and M.~P. Vecchi,
  ``\BIBforeignlanguage{en}{Optimization by {Simulated} {Annealing}},''
  \emph{\BIBforeignlanguage{en}{Science}}, vol. 220, no. 4598, pp. 671--680,
  May 1983. [Online]. Available:
  \url{https://www.science.org/doi/10.1126/science.220.4598.671}
\BIBentrySTDinterwordspacing

\bibitem{barahona1982}
\BIBentryALTinterwordspacing
F.~Barahona, ``On the computational complexity of {{Ising}} spin glass
  models,'' \emph{Journal of Physics A: Mathematical and General}, vol.~15,
  no.~10, p. 3241, 1982. [Online]. Available:
  \url{https://dx.doi.org/10.1088/0305-4470/15/10/028}
\BIBentrySTDinterwordspacing

\bibitem{ising1925}
\BIBentryALTinterwordspacing
E.~Ising, ``Beitrag zur theorie des ferromagnetismus,'' \emph{Zeitschrift für
  Physik}, vol.~31, no.~1, pp. 253--258, 1925. [Online]. Available:
  \url{https://doi.org/10.1007/BF02980577}
\BIBentrySTDinterwordspacing

\bibitem{metropolis1953}
\BIBentryALTinterwordspacing
N.~Metropolis, A.~W. Rosenbluth, M.~N. Rosenbluth, A.~H. Teller, and E.~Teller,
  ``Equation of {{State Calculations}} by {{Fast Computing Machines}},''
  \emph{The Journal of Chemical Physics}, vol.~21, no.~6, pp. 1087--1092, 1953.
  [Online]. Available: \url{https://aip.scitation.org/doi/10.1063/1.1699114}
\BIBentrySTDinterwordspacing

\bibitem{hastings1970}
\BIBentryALTinterwordspacing
W.~K. Hastings, ``Monte {{Carlo}} sampling methods using {{Markov}} chains and
  their applications,'' \emph{Biometrika}, vol.~57, no.~1, pp. 97--109, 1970.
  [Online]. Available: \url{https://doi.org/10.1093/biomet/57.1.97}
\BIBentrySTDinterwordspacing

\bibitem{gelman1997}
\BIBentryALTinterwordspacing
A.~Gelman, W.~R. Gilks, and G.~O. Roberts, ``Weak convergence and optimal
  scaling of random walk {Metropolis} algorithms,'' \emph{The Annals of Applied
  Probability}, vol.~7, no.~1, Feb. 1997, publisher: Institute of Mathematical
  Statistics. [Online]. Available:
  \url{https://doi.org/10.1214/aoap/1034625254}
\BIBentrySTDinterwordspacing

\bibitem{layden2022}
\BIBentryALTinterwordspacing
D.~Layden, G.~Mazzola, R.~V. Mishmash, M.~Motta, P.~Wocjan, J.-S. Kim, and
  S.~Sheldon. Quantum-enhanced {{Markov}} chain {{Monte Carlo}}. [Online].
  Available: \url{http://arxiv.org/abs/2203.12497}
\BIBentrySTDinterwordspacing

\bibitem{wolff1989}
\BIBentryALTinterwordspacing
U.~Wolff, ``Collective {{Monte Carlo Updating}} for {{Spin Systems}},''
  \emph{Physical Review Letters}, vol.~62, no.~4, pp. 361--364, 1989. [Online].
  Available: \url{https://link.aps.org/doi/10.1103/PhysRevLett.62.361}
\BIBentrySTDinterwordspacing

\bibitem{houdayer2001}
\BIBentryALTinterwordspacing
J.~Houdayer, ``A {{Cluster Monte Carlo Algorithm}} for 2-{{Dimensional Spin
  Glasses}},'' \emph{The European Physical Journal B}, vol.~22, no.~4, pp.
  479--484, 2001. [Online]. Available:
  \url{http://arxiv.org/abs/cond-mat/0101116}
\BIBentrySTDinterwordspacing

\bibitem{farhi2014a}
\BIBentryALTinterwordspacing
E.~Farhi, J.~Goldstone, and S.~Gutmann. A {{Quantum Approximate Optimization
  Algorithm}}. [Online]. Available: \url{http://arxiv.org/abs/1411.4028}
\BIBentrySTDinterwordspacing

\bibitem{hadfield2019}
\BIBentryALTinterwordspacing
S.~Hadfield, Z.~Wang, B.~O'Gorman, E.~G. Rieffel, D.~Venturelli, and R.~Biswas,
  ``From the {{Quantum Approximate Optimization Algorithm}} to a {{Quantum
  Alternating Operator Ansatz}},'' \emph{Algorithms}, vol.~12, no.~2, p.~34,
  2019. [Online]. Available: \url{http://arxiv.org/abs/1709.03489}
\BIBentrySTDinterwordspacing

\bibitem{carleo2017}
\BIBentryALTinterwordspacing
G.~Carleo and M.~Troyer, ``Solving the {Quantum} {Many}-{Body} {Problem} with
  {Artificial} {Neural} {Networks},'' \emph{Science}, vol. 355, no. 6325, pp.
  602--606, Feb. 2017, arXiv:1606.02318 [cond-mat, physics:quant-ph]. [Online].
  Available: \url{http://arxiv.org/abs/1606.02318}
\BIBentrySTDinterwordspacing

\bibitem{schollwoeck_2011}
\BIBentryALTinterwordspacing
U.~Schollwöck, ``The density-matrix renormalization group in the age of matrix
  product states,'' \emph{Annals of Physics}, vol. 326, no.~1, pp. 96--192, jan
  2011. [Online]. Available: \url{https://doi.org/10.1016\%2Fj.aop.2010.09.012}
\BIBentrySTDinterwordspacing

\bibitem{orus2014}
\BIBentryALTinterwordspacing
R.~Orús, ``\BIBforeignlanguage{en}{A practical introduction to tensor
  networks: {Matrix} product states and projected entangled pair states},''
  \emph{\BIBforeignlanguage{en}{Annals of Physics}}, vol. 349, pp. 117--158,
  Oct. 2014. [Online]. Available:
  \url{https://www.sciencedirect.com/science/article/pii/S0003491614001596}
\BIBentrySTDinterwordspacing

\bibitem{jia2019}
\BIBentryALTinterwordspacing
Z.-A. Jia, B.~Yi, R.~Zhai, Y.-C. Wu, G.-C. Guo, and G.-P. Guo,
  ``\BIBforeignlanguage{en}{Quantum {Neural} {Network} {States}: {A} {Brief}
  {Review} of {Methods} and {Applications}},''
  \emph{\BIBforeignlanguage{en}{Advanced Quantum Technologies}}, vol.~2, no.
  7-8, p. 1800077, 2019, \_eprint:
  https://onlinelibrary.wiley.com/doi/pdf/10.1002/qute.201800077. [Online].
  Available:
  \url{https://onlinelibrary.wiley.com/doi/abs/10.1002/qute.201800077}
\BIBentrySTDinterwordspacing

\bibitem{eisert2010}
\BIBentryALTinterwordspacing
J.~Eisert, M.~Cramer, and M.~B. Plenio, ``Colloquium: {Area} laws for the
  entanglement entropy,'' \emph{Reviews of Modern Physics}, vol.~82, no.~1, pp.
  277--306, Feb. 2010, publisher: American Physical Society. [Online].
  Available: \url{https://link.aps.org/doi/10.1103/RevModPhys.82.277}
\BIBentrySTDinterwordspacing

\bibitem{loh1990}
\BIBentryALTinterwordspacing
E.~Y. Loh, J.~E. Gubernatis, R.~T. Scalettar, S.~R. White, D.~J. Scalapino, and
  R.~L. Sugar, ``Sign problem in the numerical simulation of many-electron
  systems,'' \emph{Physical Review B}, vol.~41, no.~13, pp. 9301--9307, May
  1990, publisher: American Physical Society. [Online]. Available:
  \url{https://link.aps.org/doi/10.1103/PhysRevB.41.9301}
\BIBentrySTDinterwordspacing

\bibitem{li2019}
\BIBentryALTinterwordspacing
Z.-X. Li and H.~Yao, ``\BIBforeignlanguage{en}{Sign-{Problem}-{Free}
  {Fermionic} {Quantum} {Monte} {Carlo}: {Developments} and {Applications}},''
  \emph{\BIBforeignlanguage{en}{Annual Review of Condensed Matter Physics}},
  vol.~10, no.~1, pp. 337--356, Mar. 2019, arXiv:1805.08219 [cond-mat].
  [Online]. Available: \url{http://arxiv.org/abs/1805.08219}
\BIBentrySTDinterwordspacing

\bibitem{troyer2005}
\BIBentryALTinterwordspacing
M.~Troyer and U.-J. Wiese, ``Computational {Complexity} and {Fundamental}
  {Limitations} to {Fermionic} {Quantum} {Monte} {Carlo} {Simulations},''
  \emph{Physical Review Letters}, vol.~94, no.~17, p. 170201, May 2005,
  publisher: American Physical Society. [Online]. Available:
  \url{https://link.aps.org/doi/10.1103/PhysRevLett.94.170201}
\BIBentrySTDinterwordspacing

\bibitem{ceperley1977}
\BIBentryALTinterwordspacing
D.~Ceperley, G.~V. Chester, and M.~H. Kalos, ``Monte {Carlo} simulation of a
  many-fermion study,'' \emph{Physical Review B}, vol.~16, no.~7, pp.
  3081--3099, Oct. 1977, publisher: American Physical Society. [Online].
  Available: \url{https://link.aps.org/doi/10.1103/PhysRevB.16.3081}
\BIBentrySTDinterwordspacing

\bibitem{xia2018}
\BIBentryALTinterwordspacing
R.~Xia and S.~Kais, ``Quantum {{Machine Learning}} for {{Electronic Structure
  Calculations}},'' \emph{Nature Communications}, vol.~9, no.~1, p. 4195, 2018.
  [Online]. Available: \url{http://arxiv.org/abs/1803.10296}
\BIBentrySTDinterwordspacing

\bibitem{huggins2022}
\BIBentryALTinterwordspacing
W.~J. Huggins, B.~A. O’Gorman, N.~C. Rubin, D.~R. Reichman, R.~Babbush, and
  J.~Lee, ``\BIBforeignlanguage{en}{Unbiasing fermionic quantum {Monte} {Carlo}
  with a quantum computer},'' \emph{\BIBforeignlanguage{en}{Nature}}, vol. 603,
  no. 7901, pp. 416--420, Mar. 2022, number: 7901 Publisher: Nature Publishing
  Group. [Online]. Available:
  \url{https://www.nature.com/articles/s41586-021-04351-z}
\BIBentrySTDinterwordspacing

\bibitem{mcmillan1965}
\BIBentryALTinterwordspacing
W.~L. McMillan, ``Ground {State} of {Liquid}
  \$\{{\textbackslash}mathrm\{{He}\}\}{\textasciicircum}\{4\}\$,''
  \emph{Physical Review}, vol. 138, no.~2A, pp. A442--A451, Apr. 1965,
  publisher: American Physical Society. [Online]. Available:
  \url{https://link.aps.org/doi/10.1103/PhysRev.138.A442}
\BIBentrySTDinterwordspacing

\bibitem{sorella2007}
\BIBentryALTinterwordspacing
S.~Sorella, M.~Casula, and D.~Rocca, ``\BIBforeignlanguage{en}{Weak binding
  between two aromatic rings: {Feeling} the van der {Waals} attraction by
  quantum {Monte} {Carlo} methods},'' \emph{\BIBforeignlanguage{en}{The Journal
  of Chemical Physics}}, vol. 127, no.~1, p. 014105, Jul. 2007. [Online].
  Available: \url{https://pubs.aip.org/aip/jcp/article/906474}
\BIBentrySTDinterwordspacing

\bibitem{sorella1998}
\BIBentryALTinterwordspacing
S.~Sorella, ``Green {Function} {Monte} {Carlo} with {Stochastic}
  {Reconfiguration},'' \emph{Physical Review Letters}, vol.~80, no.~20, pp.
  4558--4561, May 1998, publisher: American Physical Society. [Online].
  Available: \url{https://link.aps.org/doi/10.1103/PhysRevLett.80.4558}
\BIBentrySTDinterwordspacing

\bibitem{choo2019}
\BIBentryALTinterwordspacing
K.~Choo, T.~Neupert, and G.~Carleo, ``Study of the {{Two-Dimensional Frustrated
  J1-J2 Model}} with {{Neural Network Quantum States}},'' \emph{Physical Review
  B}, vol. 100, no.~12, p. 125124, 2019. [Online]. Available:
  \url{http://arxiv.org/abs/1903.06713}
\BIBentrySTDinterwordspacing

\bibitem{ferrari2019}
\BIBentryALTinterwordspacing
F.~Ferrari, F.~Becca, and J.~Carrasquilla, ``Neural {Gutzwiller}-projected
  variational wave functions,'' \emph{Physical Review B}, vol. 100, no.~12, p.
  125131, Sep. 2019, publisher: American Physical Society. [Online]. Available:
  \url{https://link.aps.org/doi/10.1103/PhysRevB.100.125131}
\BIBentrySTDinterwordspacing

\bibitem{mcclean2016}
\BIBentryALTinterwordspacing
J.~R. McClean, J.~Romero, R.~Babbush, and A.~Aspuru-Guzik,
  ``\BIBforeignlanguage{en}{The theory of variational hybrid quantum-classical
  algorithms},'' \emph{\BIBforeignlanguage{en}{New Journal of Physics}},
  vol.~18, no.~2, p. 023023, Feb. 2016, publisher: IOP Publishing. [Online].
  Available: \url{https://dx.doi.org/10.1088/1367-2630/18/2/023023}
\BIBentrySTDinterwordspacing

\bibitem{tilly2022}
\BIBentryALTinterwordspacing
J.~Tilly, H.~Chen, S.~Cao, D.~Picozzi, K.~Setia, Y.~Li, E.~Grant, L.~Wossnig,
  I.~Rungger, G.~H. Booth, and J.~Tennyson, ``\BIBforeignlanguage{en}{The
  {Variational} {Quantum} {Eigensolver}: {A} review of methods and best
  practices},'' \emph{\BIBforeignlanguage{en}{Physics Reports}}, vol. 986, pp.
  1--128, Nov. 2022. [Online]. Available:
  \url{https://www.sciencedirect.com/science/article/pii/S0370157322003118}
\BIBentrySTDinterwordspacing

\bibitem{stokes2020}
\BIBentryALTinterwordspacing
J.~Stokes, J.~Izaac, N.~Killoran, and G.~Carleo, ``Quantum {Natural}
  {Gradient},'' \emph{Quantum}, vol.~4, p. 269, May 2020, arXiv:1909.02108
  [quant-ph, stat]. [Online]. Available: \url{http://arxiv.org/abs/1909.02108}
\BIBentrySTDinterwordspacing

\bibitem{hinton1986}
G.~E. Hinton and T.~J. Sejnowski, ``Learning and relearning in {Boltzmann}
  machines,'' in \emph{Parallel distributed processing: explorations in the
  microstructure of cognition, vol. 1: foundations}.\hskip 1em plus 0.5em minus
  0.4em\relax Cambridge, MA, USA: MIT Press, Jan. 1986, pp. 282--317.

\bibitem{hinton2012}
\BIBentryALTinterwordspacing
G.~E. Hinton, ``\BIBforeignlanguage{en}{A {Practical} {Guide} to {Training}
  {Restricted} {Boltzmann} {Machines}},'' in
  \emph{\BIBforeignlanguage{en}{Neural {Networks}: {Tricks} of the {Trade}:
  {Second} {Edition}}}, ser. Lecture {Notes} in {Computer} {Science},
  G.~Montavon, G.~B. Orr, and K.-R. Müller, Eds.\hskip 1em plus 0.5em minus
  0.4em\relax Berlin, Heidelberg: Springer, 2012, pp. 599--619. [Online].
  Available: \url{https://doi.org/10.1007/978-3-642-35289-8_32}
\BIBentrySTDinterwordspacing

\bibitem{hinton2006a}
G.~E. Hinton, S.~Osindero, and Y.-W. Teh, ``\BIBforeignlanguage{eng}{A fast
  learning algorithm for deep belief nets},''
  \emph{\BIBforeignlanguage{eng}{Neural Computation}}, vol.~18, no.~7, pp.
  1527--1554, Jul. 2006.

\bibitem{larochelle2012}
H.~Larochelle, M.~Mandel, R.~Pascanu, and Y.~Bengio, ``Learning algorithms for
  the classification restricted {Boltzmann} machine,'' \emph{The Journal of
  Machine Learning Research}, vol.~13, no.~1, pp. 643--669, Mar. 2012.

\bibitem{crawford2019}
\BIBentryALTinterwordspacing
D.~Crawford, A.~Levit, N.~Ghadermarzy, J.~S. Oberoi, and P.~Ronagh,
  ``Reinforcement {Learning} {Using} {Quantum} {Boltzmann} {Machines},'' Jan.
  2019, arXiv:1612.05695 [quant-ph]. [Online]. Available:
  \url{http://arxiv.org/abs/1612.05695}
\BIBentrySTDinterwordspacing

\bibitem{tang2012}
Y.~Tang, R.~Salakhutdinov, and G.~Hinton, ``Robust {Boltzmann} {Machines} for
  recognition and denoising,'' in \emph{2012 {IEEE} {Conference} on {Computer}
  {Vision} and {Pattern} {Recognition}}, Jun. 2012, pp. 2264--2271, iSSN:
  1063-6919.

\bibitem{sarker2021}
\BIBentryALTinterwordspacing
I.~H. Sarker, ``\BIBforeignlanguage{en}{Deep {Cybersecurity}: {A}
  {Comprehensive} {Overview} from {Neural} {Network} and {Deep} {Learning}
  {Perspective}},'' \emph{\BIBforeignlanguage{en}{SN Computer Science}},
  vol.~2, no.~3, p. 154, Mar. 2021. [Online]. Available:
  \url{https://doi.org/10.1007/s42979-021-00535-6}
\BIBentrySTDinterwordspacing

\bibitem{vantulder2014}
G.~van Tulder and M.~de~Bruijne, ``\BIBforeignlanguage{en}{Learning {Features}
  for {Tissue} {Classification} with the {Classification} {Restricted}
  {Boltzmann} {Machine}},'' in \emph{\BIBforeignlanguage{en}{Medical {Computer}
  {Vision}: {Algorithms} for {Big} {Data}}}, ser. Lecture {Notes} in {Computer}
  {Science}, B.~Menze, G.~Langs, A.~Montillo, M.~Kelm, H.~Müller, S.~Zhang,
  W.~T. Cai, and D.~Metaxas, Eds.\hskip 1em plus 0.5em minus 0.4em\relax Cham:
  Springer International Publishing, 2014, pp. 47--58.

\bibitem{geman1984}
S.~Geman and D.~Geman, ``Stochastic {Relaxation}, {Gibbs} {Distributions}, and
  the {Bayesian} {Restoration} of {Images},'' \emph{IEEE Transactions on
  Pattern Analysis and Machine Intelligence}, vol. PAMI-6, no.~6, pp. 721--741,
  Nov. 1984, conference Name: IEEE Transactions on Pattern Analysis and Machine
  Intelligence.

\bibitem{pmlr-vR5-carreira-perpinan05a}
\BIBentryALTinterwordspacing
M.~A. Carreira-Perpinan and G.~Hinton, ``On contrastive divergence learning,''
  in \emph{Proceedings of the tenth international workshop on artificial
  intelligence and statistics}, ser. Proceedings of machine learning research,
  R.~G. Cowell and Z.~Ghahramani, Eds., vol.~R5.\hskip 1em plus 0.5em minus
  0.4em\relax PMLR, Jan. 2005, pp. 33--40. [Online]. Available:
  \url{https://proceedings.mlr.press/r5/carreira-perpinan05a.html}
\BIBentrySTDinterwordspacing

\bibitem{montufar2014}
\BIBentryALTinterwordspacing
G.~Montufar, J.~Rauh, and N.~Ay, ``\BIBforeignlanguage{en}{Expressive {Power}
  and {Approximation} {Errors} of {Restricted} {Boltzmann} {Machines}},'' Jun.
  2014, arXiv:1406.3140 [math, stat]. [Online]. Available:
  \url{http://arxiv.org/abs/1406.3140}
\BIBentrySTDinterwordspacing

\bibitem{wiebe2015}
\BIBentryALTinterwordspacing
N.~Wiebe, A.~Kapoor, and K.~M. Svore, ``Quantum {Deep} {Learning},'' May 2015,
  arXiv:1412.3489 [quant-ph]. [Online]. Available:
  \url{http://arxiv.org/abs/1412.3489}
\BIBentrySTDinterwordspacing

\bibitem{zoufal2021}
\BIBentryALTinterwordspacing
C.~Zoufal, A.~Lucchi, and S.~Woerner, ``\BIBforeignlanguage{en}{Variational
  quantum {Boltzmann} machines},'' \emph{\BIBforeignlanguage{en}{Quantum
  Machine Intelligence}}, vol.~3, no.~1, p.~7, Feb. 2021. [Online]. Available:
  \url{https://doi.org/10.1007/s42484-020-00033-7}
\BIBentrySTDinterwordspacing

\bibitem{amin2015}
\BIBentryALTinterwordspacing
M.~Amin, ``Searching for quantum speedup in quasistatic quantum annealers,''
  \emph{Physical Review A}, vol.~92, no.~5, p. 052323, Nov. 2015,
  arXiv:1503.04216 [cond-mat, physics:quant-ph]. [Online]. Available:
  \url{http://arxiv.org/abs/1503.04216}
\BIBentrySTDinterwordspacing

\bibitem{benedetti2016}
\BIBentryALTinterwordspacing
M.~Benedetti, J.~Realpe-Gómez, R.~Biswas, and A.~Perdomo-Ortiz, ``Estimation
  of effective temperatures in quantum annealers for sampling applications: {A}
  case study with possible applications in deep learning,'' \emph{Physical
  Review A}, vol.~94, no.~2, p. 022308, Aug. 2016, arXiv:1510.07611 [quant-ph].
  [Online]. Available: \url{http://arxiv.org/abs/1510.07611}
\BIBentrySTDinterwordspacing

\bibitem{adachi2015}
\BIBentryALTinterwordspacing
S.~H. Adachi and M.~P. Henderson, ``Application of {Quantum} {Annealing} to
  {Training} of {Deep} {Neural} {Networks},'' Oct. 2015, arXiv:1510.06356
  [quant-ph, stat]. [Online]. Available: \url{http://arxiv.org/abs/1510.06356}
\BIBentrySTDinterwordspacing

\bibitem{sajjan2021}
\BIBentryALTinterwordspacing
M.~Sajjan, S.~H. Sureshbabu, and S.~Kais, ``Quantum {Machine}-{Learning} for
  {Eigenstate} {Filtration} in {Two}-{Dimensional} {Materials},'' \emph{Journal
  of the American Chemical Society}, vol. 143, no.~44, pp. 18\,426--18\,445,
  Nov. 2021, arXiv:2105.09488 [physics]. [Online]. Available:
  \url{http://arxiv.org/abs/2105.09488}
\BIBentrySTDinterwordspacing

\bibitem{sureshbabu2021}
\BIBentryALTinterwordspacing
S.~H. Sureshbabu, M.~Sajjan, S.~Oh, and S.~Kais, ``Implementation of {Quantum}
  {Machine} {Learning} for {Electronic} {Structure} {Calculations} of
  {Periodic} {Systems} on {Quantum} {Computing} {Devices},'' \emph{Journal of
  Chemical Information and Modeling}, vol.~61, no.~6, pp. 2667--2674, Jun.
  2021, publisher: American Chemical Society. [Online]. Available:
  \url{https://doi.org/10.1021/acs.jcim.1c00294}
\BIBentrySTDinterwordspacing

\bibitem{paul2023}
\BIBentryALTinterwordspacing
F.~Paul, M.~Falkenthal, and S.~Feld, ``Clever {Design}, {Unexpected}
  {Obstacles}: {Insights} on {Implementing} a {Quantum} {Boltzmann}
  {Machine},'' Jan. 2023, arXiv:2301.13705 [quant-ph]. [Online]. Available:
  \url{http://arxiv.org/abs/2301.13705}
\BIBentrySTDinterwordspacing

\bibitem{bukov2021}
\BIBentryALTinterwordspacing
M.~Bukov, M.~Schmitt, and M.~Dupont, ``Learning the ground state of a
  non-stoquastic quantum {Hamiltonian} in a rugged neural network landscape,''
  \emph{SciPost Physics}, vol.~10, no.~6, p. 147, Jun. 2021, arXiv:2011.11214
  [cond-mat, physics:physics, physics:quant-ph]. [Online]. Available:
  \url{http://arxiv.org/abs/2011.11214}
\BIBentrySTDinterwordspacing

\bibitem{park2022}
\BIBentryALTinterwordspacing
C.-Y. Park and M.~J. Kastoryano, ``Expressive power of complex-valued
  restricted {Boltzmann} machines for solving non-stoquastic {Hamiltonians},''
  \emph{Physical Review B}, vol. 106, no.~13, p. 134437, Oct. 2022,
  arXiv:2012.08889 [cond-mat, physics:quant-ph]. [Online]. Available:
  \url{http://arxiv.org/abs/2012.08889}
\BIBentrySTDinterwordspacing

\bibitem{westerhout2020}
T.~Westerhout, N.~Astrakhantsev, K.~S. Tikhonov, M.~I. Katsnelson, and A.~A.
  Bagrov, ``\BIBforeignlanguage{eng}{Generalization properties of neural
  network approximations to frustrated magnet ground states},''
  \emph{\BIBforeignlanguage{eng}{Nature Communications}}, vol.~11, no.~1, p.
  1593, Mar. 2020.

\bibitem{amin2018}
\BIBentryALTinterwordspacing
M.~H. Amin, E.~Andriyash, J.~Rolfe, B.~Kulchytskyy, and R.~Melko,
  ``\BIBforeignlanguage{en}{Quantum {Boltzmann} {Machine}},''
  \emph{\BIBforeignlanguage{en}{Physical Review X}}, vol.~8, no.~2, p. 021050,
  May 2018, arXiv:1601.02036 [quant-ph]. [Online]. Available:
  \url{http://arxiv.org/abs/1601.02036}
\BIBentrySTDinterwordspacing

\bibitem{Benedetti17}
M.~Benedetti, J.~Realpe-G{\'o}mez, R.~Biswas, and A.~Perdomo-Ortiz,
  ``Quantum-assisted learning of hardware-embedded probabilistic graphical
  models,'' \emph{Physical Review X}, vol.~7, no.~4, p. 041052, 2017.

\bibitem{stephenson_introduction_2000}
T.~A. Stephenson, Ed., \emph{An {Introduction} to {Bayesian} {Network} {Theory}
  and {Usage}}.\hskip 1em plus 0.5em minus 0.4em\relax IDIAP, 2000.

\bibitem{drury_survey_2017}
\BIBentryALTinterwordspacing
B.~Drury, J.~Valverde-Rebaza, M.-F. Moura, and A.~de~Andrade~Lopes, ``A survey
  of the applications of {Bayesian} networks in agriculture,''
  \emph{Engineering Applications of Artificial Intelligence}, vol.~65, pp.
  29--42, Oct. 2017. [Online]. Available:
  \url{https://www.sciencedirect.com/science/article/pii/S0952197617301513}
\BIBentrySTDinterwordspacing

\bibitem{heckerman_bayesian_1997}
\BIBentryALTinterwordspacing
D.~Heckerman, ``Bayesian {Networks} for {Data} {Mining},'' \emph{Data Mining
  and Knowledge Discovery}, vol.~1, no.~1, pp. 79--119, Mar. 1997. [Online].
  Available: \url{https://doi.org/10.1023/A:1009730122752}
\BIBentrySTDinterwordspacing

\bibitem{cano_applications_2004}
\BIBentryALTinterwordspacing
R.~Cano, C.~Sordo, and J.~M. Gutiérrez, ``Applications of {Bayesian}
  {Networks} in {Meteorology},'' in \emph{Advances in {Bayesian} {Networks}},
  ser. Studies in {Fuzziness} and {Soft} {Computing}, J.~A. Gámez, S.~Moral,
  and A.~Salmerón, Eds.\hskip 1em plus 0.5em minus 0.4em\relax Berlin,
  Heidelberg: Springer, 2004, pp. 309--328. [Online]. Available:
  \url{https://doi.org/10.1007/978-3-540-39879-0_17}
\BIBentrySTDinterwordspacing

\bibitem{tosun_systematic_2017}
\BIBentryALTinterwordspacing
A.~Tosun, A.~B. Bener, and S.~Akbarinasaji, ``A systematic literature review on
  the applications of {Bayesian} networks to predict software quality,''
  \emph{Software Quality Journal}, vol.~25, no.~1, pp. 273--305, Mar. 2017.
  [Online]. Available: \url{https://doi.org/10.1007/s11219-015-9297-z}
\BIBentrySTDinterwordspacing

\bibitem{chickering_learning_1996}
\BIBentryALTinterwordspacing
D.~M. Chickering, ``Learning {Bayesian} {Networks} is {NP}-{Complete},'' in
  \emph{Learning from {Data}: {Artificial} {Intelligence} and {Statistics}
  {V}}, ser. Lecture {Notes} in {Statistics}, D.~Fisher and H.-J. Lenz,
  Eds.\hskip 1em plus 0.5em minus 0.4em\relax New York, NY: Springer, 1996, pp.
  121--130. [Online]. Available:
  \url{https://doi.org/10.1007/978-1-4612-2404-4_12}
\BIBentrySTDinterwordspacing

\bibitem{koller_probabilistic_2009}
D.~Koller and N.~Friedman, \emph{Probabilistic {Graphical} {Models}:
  {Principles} and {Techniques}}.\hskip 1em plus 0.5em minus 0.4em\relax MIT
  Press, Jul. 2009, google-Books-ID: 7dzpHCHzNQ4C.

\bibitem{low_quantum_2014}
\BIBentryALTinterwordspacing
G.~H. Low, T.~J. Yoder, and I.~L. Chuang, ``Quantum inference on {Bayesian}
  networks,'' \emph{Physical Review A}, vol.~89, no.~6, p. 062315, Jun. 2014,
  publisher: American Physical Society. [Online]. Available:
  \url{https://link.aps.org/doi/10.1103/PhysRevA.89.062315}
\BIBentrySTDinterwordspacing

\bibitem{gao_enhancing_2022}
\BIBentryALTinterwordspacing
X.~Gao, E.~R. Anschuetz, S.-T. Wang, J.~I. Cirac, and M.~D. Lukin, ``Enhancing
  {Generative} {Models} via {Quantum} {Correlations},'' \emph{Physical Review
  X}, vol.~12, no.~2, p. 021037, May 2022, publisher: American Physical
  Society. [Online]. Available:
  \url{https://link.aps.org/doi/10.1103/PhysRevX.12.021037}
\BIBentrySTDinterwordspacing

\bibitem{zeng_quantum_2016}
\BIBentryALTinterwordspacing
W.~Zeng and B.~Coecke, ``\BIBforeignlanguage{en}{Quantum {Algorithms} for
  {Compositional} {Natural} {Language} {Processing}},''
  \emph{\BIBforeignlanguage{en}{Electronic Proceedings in Theoretical Computer
  Science}}, vol. 221, pp. 67--75, Aug. 2016. [Online]. Available:
  \url{http://arxiv.org/abs/1608.01406}
\BIBentrySTDinterwordspacing

\bibitem{meichanetzidis_grammar-aware_2023}
\BIBentryALTinterwordspacing
K.~Meichanetzidis, A.~Toumi, G.~de~Felice, and B.~Coecke,
  ``\BIBforeignlanguage{en}{Grammar-aware sentence classification on quantum
  computers},'' \emph{\BIBforeignlanguage{en}{Quantum Machine Intelligence}},
  vol.~5, no.~1, p.~10, Feb. 2023. [Online]. Available:
  \url{https://doi.org/10.1007/s42484-023-00097-1}
\BIBentrySTDinterwordspacing

\bibitem{coecke_mathematical_2010}
\BIBentryALTinterwordspacing
B.~Coecke, M.~Sadrzadeh, and S.~Clark, ``Mathematical {Foundations} for a
  {Compositional} {Distributional} {Model} of {Meaning},'' Mar. 2010,
  arXiv:1003.4394 [cs, math]. [Online]. Available:
  \url{http://arxiv.org/abs/1003.4394}
\BIBentrySTDinterwordspacing

\bibitem{needham_inference_2006}
C.~J. Needham, J.~R. Bradford, A.~J. Bulpitt, and D.~R. Westhead, ``Inference
  in {Bayesian} networks,'' \emph{Nature biotechnology}, vol.~24, no.~1, pp.
  51--53, 2006, publisher: Nature Publishing Group US New York.

\bibitem{liu_application_2010}
Y.~Liu and J.-D.~J. Han, ``Application of {Bayesian} networks on large-scale
  biological data,'' \emph{Frontiers in Biology}, vol.~5, pp. 98--104, 2010,
  publisher: Springer.

\bibitem{schaapveld_bayesian_2019}
T.~E. Schaapveld, S.~L. Opperman, and S.~Harbison, ``Bayesian networks for the
  interpretation of biological evidence,'' \emph{Wiley Interdisciplinary
  Reviews: Forensic Science}, vol.~1, no.~3, p. e1325, 2019, publisher: Wiley
  Online Library.

\bibitem{lau_bayesian_2016}
C.~L. Lau and C.~S. Smith, ``Bayesian networks in infectious disease
  eco-epidemiology,'' \emph{Reviews on environmental health}, vol.~31, no.~1,
  pp. 173--177, 2016, publisher: De Gruyter.

\bibitem{seixas_bayesian_2014}
F.~L. Seixas, B.~Zadrozny, J.~Laks, A.~Conci, and D.~C.~M. Saade, ``A
  {Bayesian} network decision model for supporting the diagnosis of dementia,
  {Alzheimer}'s disease and mild cognitive impairment,'' \emph{Computers in
  biology and medicine}, vol.~51, pp. 140--158, 2014, publisher: Elsevier.

\bibitem{bielza_bayesian_2014}
C.~Bielza and P.~Larrañaga, ``Bayesian networks in neuroscience: a survey,''
  \emph{Frontiers in computational neuroscience}, vol.~8, p. 131, 2014,
  publisher: Frontiers Media SA.

\bibitem{sebastiani_bayesian_2005}
P.~Sebastiani, M.~Abad, and M.~F. Ramoni, ``Bayesian networks for genomic
  analysis,'' \emph{Genomic signal processing and statistics}, vol.~2, pp.
  281--320, 2005, publisher: EURASIP Book Series on Signal Processing and
  Communications, Hindawi ….

\bibitem{murphy_dynamic_2002}
K.~P. Murphy, ``Dynamic bayesian networks,'' \emph{Probabilistic Graphical
  Models, M. Jordan}, vol.~7, p. 431, 2002.

\bibitem{ghahramani_learning_2006}
Z.~Ghahramani, ``Learning dynamic {Bayesian} networks,'' \emph{Adaptive
  Processing of Sequences and Data Structures: International Summer School on
  Neural Networks “ER Caianiello” Vietri sul Mare, Salerno, Italy September
  6–13, 1997 Tutorial Lectures}, pp. 168--197, 2006, publisher: Springer.

\bibitem{song_time-varying_2009}
L.~Song, M.~Kolar, and E.~Xing, ``Time-varying dynamic bayesian networks,''
  \emph{Advances in neural information processing systems}, vol.~22, 2009.

\end{thebibliography}
